\documentclass[a4paper,10pt]{article}
\usepackage[utf8]{inputenc}
\usepackage{amsmath}
\usepackage{amsfonts}
\usepackage{epsfig}
\usepackage{graphicx}
\usepackage{cite}
\title{Integrable model of bosons in a four-well ring with anisotropic tunneling}
\author{A P Tonel and L H Ymai \\
Universidade Federal do Pampa, \\
Travessa 45, n 1650, \\
Bairro Malafaia, Bag\'e, RS, Brazil \\
email: arlei.tonel@unipampa.edu.br, \\
leandro.ymai@unipampa.edu.br
  \\
  ~~\\
A Foerster \\
Instituto de F\'isica da UFRGS, \\
Av. Bento Gon\c{c}alves 9500, \\
Agronomia, Porto Alegre, RS, Brazil
    \\
email: angela@if.ufrgs.br  \\
~~\\   
J Links\\
School of Mathematics and Physics, \\
The University of Queensland, \\ Brisbane, QLD 4072, Australia \\
email: jrl@maths.uq.edu.au}
\date{}
\begin{document}

\maketitle

\begin{abstract}
We introduce an integrable, four-well ring model for bosons where the tunneling couplings between nearest-neighbour wells are not restricted to be equal. We show how the model may be derived through the Quantum Inverse Scattering Method from a solution of the Yang--Baxter equation, and in turn solved by algebraic Bethe Ansatz means. The model admits multiple pseudovaccum states. Numerical evidence is provided to indicate that all pseudovacua are required to obtain a complete set of Bethe eigenstates. The model has the notable property that there is a class of eigenstates which admit a simple, closed-form energy expression. 
\end{abstract}

\newpage 
\section{Introduction}

Systems of Bose-Einstein Condensates (BECs) continue to offer rich opportunities for exploring intrinsically quantum characteristics, such as superposition and entanglement,  at the mesoscopic/macroscopic interface. Models based on few bosonic modes have been widely studied. Initially, a two-mode model of quantum tunneling between two wells, specifically the two-site Bose-Hubbard model, was sufficient to predict self-trapping phenomena \cite{milburn,smerzi,leggett,tlf05} which was subsequently confirmed experimentally \cite{albiez}. Ensuing studies were extended to three-mode models to investigate, for example, dynamics \cite{nhmm01,wyw09}, entanglement, \cite{lw06},  dissipation \cite{sm10}, quantum phase transitions \cite{pwqbd14}, and condensate fragmentation \cite{fragment}.

The dynamics of a four-well Bose-Hubbard model was studied in \cite{liberato} as a means to achieve mass transport and persistent currents around a loop. The same model, but with two significantly different tunneling rates, has been investigated 
in relation to a mesoscopic quantum system in thermal contact \cite{anglin}. The quantum dynamics for a range of different initial conditions, 
in terms of the particle number distribution among the wells and quantum statistics, was presented in \cite{olsen}. 
In \cite{laha} it is claimed that an appropriate control of short-range and dipolar 
interaction may lead to the dynamical creation of mesoscopic quantum 
superposition, which could be employed in the design of Heisenberg-limited atom interferometers through a four-well system.
Also, it has been proposed to implement a two-well 
model with two levels in each well (yielding a four-mode model)  as a means to experimentally measure EPR entanglement \cite{he}.

The two-mode Bose-Hubbard model is an example of an integrable system which admits an exact Bethe Ansatz solution \cite{esks91,esse92,eks93}, the existence of which allows analytic computation of physical quantities. 
For example, a Bethe Ansatz solution of the model was used to calculate the ground-state one-body density matrix in the attractive regime \cite{lm15}. This calculation shows the existence of a quantum phase transition point of the model which is characterized as separating condensate fragmentation from an unfragmented phase. It is desirable from this perspective to also identify integrable cases of similar models with higher number of modes. This is the objective of our work here for the case of a four-well model.

In Sect. 2 we define the model and set up the notational conventions. Sect. 3 is devoted to deriving the model via the Quantum Inverse Scattering Method, and deriving an exact solution through the algebraic Bethe Ansatz \cite{fa1,kul}. Here we encounter some unusual properties for this model. One is that although we can identify four conserved operators for the system, only two of them are obtained through the transfer matrix produced by the Quantum Inverse Scattering Method. Another curiosity is that there are multiple pseudovacuum states available for the purposes of implementing the algebraic Bethe Ansatz. A consequence of this is that there is a class of eigenstates which admit a simple, closed-form energy expression. Numerical studies undertaken in Sect. 4 indicate that all pseudovacua are required in order to obtain a complete set of Bethe eigenstates. Concluding comments are given in Sect. 5.     

\section{Hamiltonian of a four-well ring with anisotropic tunneling}
The  four-well model we present has the Hamiltonian 
\begin{align}
\label{h4}
H%&=& U(N_{1,3}-N_{2,4})^2+\mu(N_{1,3}-N_{2,4})+t(A_\alpha A_\beta^\dagger+A_\beta A_\alpha^\dagger),\nonumber\\
&= U(N_1+N_3-N_2-N_4)^2+\mu(N_1+N_3-N_2-N_4)\nonumber\\
& \quad +t_{12}(a_1 a_2^\dagger+a_1^\dagger a_2)+t_{14}( a_1 a_4^\dagger+a_1^\dagger a_4)\\
& \quad +t_{23}(a_3 a_2^\dagger+a_3^\dagger a_2)+t_{34}(a_3 a_4^\dagger+a_3^\dagger a_4).\nonumber
\end{align} 
The operators  $a_i^\dagger, a_i, \;i=1,\ldots,4$ denote the single-particle creation and  
annihilation operators acting in the four wells, with  $N_i = a_i^\dagger a_i$ providing the number 
operator for bosons in well $i$. The quantities $t_{ij}$  which determine the tunneling couplings are not independent but take a factorised form $t_{ij}=-\kappa \alpha_i\alpha_j$ where the set $\{\kappa, \alpha_j,\,j=1,2,3,4\}$ are arbitrary real variables.  This is equivalent to the constraint
\begin{align*}
t_{12}t_{34}=t_{23}t_{14}
\end{align*}
but still admits sufficient freedom to investigate a range of anisotropic tunneling regimes.  
The coupling parameters $U$, which controls on-site and inter-well interactions between bosons, 
and $ \mu$, which is an external potential parameter, are also arbitrary real variables. 
 
\begin{figure}[ht]
%{\includegraphics[width=5cm]{4well.eps}}
\begin{center}
\epsfig{file=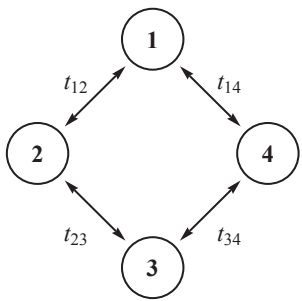,width=4.0cm,height=4cm,angle=0}
\caption{Schematic representation of the model (\ref{h4}) showing the tunneling couplings among the wells.}
\label{esquema}
\end{center}
\end{figure}

The total number operator 
$N=\sum_{i=1}^4N_i$ is a conserved operator commuting with the Hamiltonian. For each fixed value of $N$ the dimension of the associated Hilbert space $V_N$ is 
$${\rm dim}(V_N)=\frac{(N+3)!}{3!N!}=\frac{1}{6}(N+3)(N+2)(N+1). $$ 
Later will show that there are two additional constants of the motion, which leads to the conclusion that the Hamiltonian is integrable.  
In Fig. \ref{esquema} we show a schematic representation for the model.
%\vspace{0.5cm}

Physically, the Hamiltonian (\ref{h4}) models anisotropic Josephson 
tunneling for a Bose-Einstein condensate confined in a four-well ring potential. 
This Hamiltonian differs from the usual Bose-Hubbard model \cite{anglin,olsen} by the inclusion of interactions terms of the form $N_i N_j$. However these quadratic interactions are found in other models, e.g. \cite{sm10,laha}.  The inclusion of the external potential term $\mu$, which has an analogue in the case of the two-mode Bose-Hubbard model \cite{tlf205,dhc07,jmp10}, is compatible with the integrability. 

There are rich mathematical structures associated to this model, including an exact Bethe Ansatz solution which will be derived below.

\section{Quantum Inverse Scattering Method}
Our first objective is to establish that the Hamiltonian (\ref{h4}) can be obtained through the Quantum Inverse Scattering Method. We follow the general approach of \cite{gilberto}, which itself extends ideas presented in \cite{eric}. We begin with the standard $su(2)$-invariant $R$-matrix, depending on the
spectral parameter $u$:
\begin{equation}
R(u) =  \left ( \begin {array} {cccc}
1&0&0&0\\
0&b(u)&c(u)&0\\
0&c(u)&b(u)&0\\
0&0&0&1\\
\end {array} \right ),
\label{r}
\end{equation}
with $b(u)=u/(u+\eta)$ and $c(u)=\eta/(u+\eta)$.
Above, $\eta$ is a free real parameter.
It is well-known that $R(u)$ satisfies the Yang--Baxter equation \cite{y67,b72}
\begin{equation}
R _{12} (u-v)  R _{13} (u)  R _{23} (v) =
R _{23} (v)  R _{13}(u)  R _{12} (u-v).
\label{ybe}
\end{equation}
Here $R_{jk}(u)$ denotes the matrix  acting non-trivially on the
$j$-th and $k$-th spaces and as the identity on the remaining space.
We start with the Lax operator  \cite{gilberto}
\begin{equation}
L^{(i,j)}(u)=\left(\begin{matrix}
                       uI+\eta N_{i,j}& A_{i,j}\cr
                        A_{i,j}^{\dagger} & \eta^{-1}(\alpha_i^2+\alpha_j^2)I\cr
                      \end{matrix}\right),
\end{equation}
where $I$ denotes the identity operator and 
$$
N_{i,j}=N_i+N_j,\;\;\;A_{i,j}=\alpha_ia_i+\alpha_ja_j, \;\;\;A_{i,j}^{\dagger}=\alpha_ia_i^{\dagger}+\alpha_ja_j^{\dagger}.
$$
Since $L^{(i,j)}(u)$ satisfies the Yang-Baxter equation 
\begin{equation}
R_{12}(u-v)L_1^{(i,j)}(u)L_2^{(i,j)}(v)=L_2^{(i,j)}(v)L^{(i,j)}_1(u)R_{12}(u-v) ,
\end{equation}
we can define a monodromy matrix from the Yang-Baxter realization presented above as
\begin{equation}
T(u) = L^{(1,3)}(u+\omega)L^{(2,4)}(u-\omega)=
\left(\begin{matrix}
A(u)&B(u)\cr
C(u)&D(u)\cr
\end{matrix} \right) ,
\end{equation}
where 
\begin{align*}
 A(u) &=  ((u+\omega)I+\eta N_{1,3})((u-\omega)I+\eta N_{2,4})+A^{\phantom{\dagger}}_{1,3} A_{2,4}^{\dagger}   \\
 B(u) &=  ((u+\omega)I+\eta N_{1,3})A_{2,4}+\eta^{-1}(\alpha_2^2+\alpha_4^2)A_{1,3}  \\
 C(u) &= A_{1,3}^{\dagger}((u-\omega)I+\eta N_{2,4})+\eta^{-1}(\alpha_1^2+\alpha_3^2)A_{2,4}^{\dagger}  \\ 
 D(u) &=  A_{1,3}^{\dagger}A^{\phantom{\dagger}}_{2,4}+\eta^{-2}(\alpha_1^2+\alpha_3^2)(\alpha_2^2+\alpha_4^2)I.                
\end{align*}
It follows that the monodromy matrix satisfies the Yang-Baxter 
equation
\begin{equation}
R_{12}(u-v) T_1(u) T_2(v)=
T_2(v) T_1(u)R_{12}(u-v).
\label{yba}
\end{equation}
Finally, we define the transfer matrix 
\begin{equation}
\tau(u) ={\rm trace}(T(u)) = A(u)+D(u)=c_0+c_1u+c_2u^2,
\label{tm}
\end{equation}
where 
\begin{align*}
c_0&=\tau(0)=-\kappa^{-1}{H}+\left(\eta^{-2}(\alpha_1^2+\alpha_3^2)(\alpha_2^2+\alpha_4^2)+\frac{\eta^2N^2}{4}-\omega^2\right)I, \\
c_1&=\frac{d}{du}\tau(u)|_{u=0}=\eta N,  \\
c_2&= \frac{1}{2}\frac{d^2}{du^2}\tau(u)|_{u=0}=I 
\end{align*}
and the following identification has been made for the coupling constants:
\begin{align}
U=\frac{\kappa\eta^2}{4}, \qquad \mu=\kappa\omega\eta\qquad t_{ij}=-\kappa\alpha_i\alpha_j.
\label{identify}
\end{align}

It follows from (\ref{yba}) that the
transfer matrix commutes for different values of the spectral parameter,
$$[\tau(u),\tau(v)]=0,$$ 
and consequently the eigenvectors of $\tau(u)$ are independent of $u$. 
It is now straightforward to check that the Hamiltonian (\ref{h4}) is related to the 
transfer matrix  $\tau(u)$ (\ref{tm}) through
\begin{align*}
H = -\kappa\left(\tau(u)+\left(\omega^2-u^2-\eta^{-2}(\alpha_1^2+\alpha_3^2)(\alpha_2^2+\alpha_4^2)-u \eta N-\frac{\eta^2 N}{4}\right)I\right).
\end{align*}
Therefore the energy spectrum is given by
\begin{align}
E=-\kappa\left(\lambda(u)+\omega^2-u^2-\eta^{-2}(\alpha_1^2+\alpha_3^2)(\alpha_2^2+\alpha_4^2)-u \eta N-\frac{\eta^2N^2}{4}\right)
\label{nrg}
\end{align}
where $\lambda(u) $ denotes the eigenvalues of the transfer matrix. 
To find expressions for these eigenvalues we apply the algebraic Bethe Ansatz procedure \cite{fa1,kul}.

\subsection{Pseudovacua and algebraic Bethe Ansatz}
 To apply the algebraic Bethe Ansatz method we have to determine suitable pseudovacua. 
 To accomplish this we define the commuting operators
\begin{align*}
\Gamma_{1,3}^{\dagger}&=\alpha_3 a_1^\dagger-\alpha_1 a_3^\dagger,\nonumber\\
\Gamma_{2,4}^{\dagger}&=\alpha_4 a_2^\dagger-\alpha_2 a_4^\dagger,\nonumber
\end{align*}
which satisfy
\begin{align*}
\lbrack \Gamma_{1,3}^{\dagger},A_{1,3} \rbrack&=0,\ \qquad &\lbrack \Gamma_{2,4}^{\dagger},A_{2,4} \rbrack &=0,\\
\lbrack \Gamma_{1,3}^{\dagger},A_{2,4} \rbrack &=0,\ \qquad &\lbrack \Gamma_{2,4}^{\dagger},A_{1,3} \rbrack &=0,\\
\lbrack N_{1,3},(\Gamma_{1,3}^{\dagger})^l \rbrack&= l(\Gamma_{1,3}^{\dagger})^l, \qquad &\lbrack N_{2,4},(\Gamma_{2,4}^{\dagger})^k \rbrack &= k(\Gamma_{2,4}^{\dagger})^k, \\
\lbrack\Gamma_{1,3},B(u)\rbrack &=\eta\Gamma_{1,3}A_{2,4}  , \qquad &\lbrack\Gamma_{2,4},B(u)\rbrack &= 0,\\
\lbrack\Gamma^\dagger_{1,3},B(u)\rbrack&= -\eta\Gamma^\dagger_{1,3}A_{2,4} , \qquad &\lbrack\Gamma^\dagger_{2,4},B(u)\rbrack&=0,\\
\lbrack\Gamma_{1,3},C(u)\rbrack &=0, \qquad &\lbrack\Gamma_{2,4},C(u)\rbrack &= \eta \Gamma_{2,4} A_{1,3}^\dagger,\\
\lbrack\Gamma^\dagger_{1,3},C(u)\rbrack&=0 , \qquad &\lbrack\Gamma^\dagger_{2,4},C(u)\rbrack&=-\eta \Gamma^\dagger _{2,4} A_{1,3}^\dagger.
\end{align*}

Let $|0\rangle$ denote the bosonic vacuum state defined by the property
\begin{align*}
a_j|0\rangle=0 \qquad j=1,2,3,4.
\end{align*}
For the purposes of the algebraic Bethe Ansatz calculation the pseudovacua are given by 
\begin{equation}
|\phi_{k,l}\rangle = (\Gamma_{1,3}^{\dagger})^k(\Gamma_{2,4}^{\dagger})^l|0\rangle,\;\;\; k+l\leq N 
\end{equation}
and satisfy
\begin{align*}
A(u)|\phi_{k,l}\rangle &= (u+\omega+k\eta)(u-\omega+l\eta) |\phi_{k,l}\rangle, \\
B(u)|\phi_{k,l}\rangle &= 0, \\
%C(u)|\phi_{k,l}\rangle &\neq 0, \\
D(u)|\phi_{k,l}\rangle &= \eta^{-2}(\alpha_1^2+\alpha_3^2)(\alpha_2^2+\alpha_4^2)|\phi_{k,l}\rangle. \nonumber
\end{align*}
Now we can define the Bethe states associated to each pseudovacuum as 
\begin{align*}
|\psi_{k,l}\rangle=\left\{\begin{array}{ll}\displaystyle \prod_{i=1}^{N-k-l}C(u_i)|\phi_{k,l}\rangle, & \mathrm{if} \qquad k,l=0,1,\cdots,N-1, \qquad k+l< N\\
|\phi_{k,l}\rangle, &\mathrm{if} \qquad k+l=N.
\end{array}
\right.
\end{align*}
Using the following algebraic relations
\begin{align*}
A(u)C(v)&=\frac{u-v+\eta}{u-v}C(v)A(u)-\frac{\eta}{u-v}C(u)A(v),\\
D(u)C(u)&=\frac{u-v-\eta}{u-v}C(v)D(u)+\frac{\eta}{u-v}C(u)D(v)
\end{align*}
which are determined from (\ref{yba}), the algebraic Bethe Ansatz \cite{fa1,kul} can be applied. In particular it is found that 
\begin{equation}
 \tau(u)|\psi_{k,l}\rangle =\lambda_{k,l}(u)|\psi_{k,l}\rangle,
\end{equation}
where for $k+l=N$ the transfer matrix eigenvalues are given by
\begin{align*}
\lambda_{k,l}(u) &= (u+\omega+k\eta)(u-\omega+l\eta)
+\eta^{-2}(\alpha_1^2+\alpha_3^2)(\alpha_2^2+\alpha_4^2), 
\end{align*} 
while for $k+l<N$ the eigenvalues are given by
\begin{align*}
\lambda_{k,l}(u) &= (u+\omega+k\eta)(u-\omega+l\eta)\prod_{j=1}^{N-k-l}\frac{u-v_j+\eta}{u-v_j} \\
&\qquad +\eta^{-2}(\alpha_1^2+\alpha_3^2)(\alpha_2^2+\alpha_4^2)\prod_{j=1}^{N-k-l}\frac{u-v_j-\eta}{u-v_j} 
\end{align*} 
and the parameters $v_j$ are solutions of the Bethe Ansatz equations
\begin{align}
\frac{\eta^2(v_i+\omega+k\eta)(v_i-\omega+l\eta)}{(\alpha_1^2+\alpha_3^2)(\alpha_2^2+\alpha_4^2)}
=\prod_{j\neq i}^{N-k-l}\frac{v_i-v_j-\eta}{v_i-v_j+\eta}.
\label{bae} 
\end{align}
These eigenvalue formulae allow for the energy spectrum to be computed through (\ref{nrg}). We remark that in the case when $k+l=N$ the expression (\ref{nrg}) reduces to the simple form
\begin{align}
E&=-\kappa\left(kl\eta^2+\omega\eta(l-k)-\frac{\eta^2 N^2}{4}\right) \nonumber \\
&=U(l-k)^2+(l-k)\mu.
\label{nrg1}
\end{align}
Eq. (\ref{nrg1}) is notable for two reasons. First it shows that there is a class of eigenstates which admit a simple, closed-form energy eigenvalue. The second is that for these eigenstates the energy eigenvalue is independent of the variables $t_{ij}$. 

\subsection{Additional conserved operators}
The Hamiltonian (\ref{h4}) is a model with four bosonic modes, for which it is expected that integrability requires at least 
four independent conserved operators. However, the Quantum Inverse Scattering Method applied above shows that we obtain only two independent 
conserved operators, $H$ and $N$, from the transfer matrix.  Nonetheless there exists two other independent conserved operators with the form
\begin{align*}
 Q_{1,3} &= \frac{1}{\alpha_1^2+\alpha_3^2}\Gamma_{1,3}^{\dagger}\Gamma^{\phantom{\dagger}}_{1,3}\nonumber \\
 Q_{2,4} &= \frac{1}{\alpha_2^2+\alpha_4^2}\Gamma_{2,4}^{\dagger}\Gamma_{2,4}^{\phantom{\dagger}}.
\end{align*}
These operators satisfy the commutation relations
\begin{align*}
[Q_{1,3}, Q_{2,4}]=0,\;\;\;\;[H,Q_{1,3}]=[H,Q_{2,4}]=[N,Q_{1,3}]=[N,Q_{2,4}]=0
\end{align*}
so $Q_{1,3}$ and $Q_{2,4}$ together with the Hamiltonian $H$ and the total boson number operator $N$ provide 
four independent conserved operators 
for the model. Moreover, the additional conserved operators satisfy the following commutation relations
\begin{align*}
 [Q_{1,3}, C(u)]=[Q_{2,4}, C(u)]=0.
\end{align*}
It follows that each Bethe state $|\psi_{k,l}\rangle$ as defined above is a simultaneous eigenstate of the additional conserved operators:
\begin{align}
  Q_{1,3}|\psi_{k,l}\rangle &= k|\psi_{k,l}\rangle, \label{cons1}\\
  Q_{2,4}|\psi_{k,l}\rangle &= l|\psi_{k,l}\rangle. \label{cons2}
\end{align}

\section{Numerical results for small number of bosons} 

Finally we undertake a numerical analysis to demonstrate that (at least in some instances) utilizing all possible pseudovacua in the Bethe Ansatz procedure allows for a complete set of energy levels to be obtained. 

\subsection{$N =1 $}

By using the normalized Fock basis  
$$\{|\chi_j\rangle=a_j^\dagger|0\rangle: 1\leq j \leq 4\}$$ 
for the sector $N=1$, 
the Hamiltonian takes the matrix form
\begin{equation}
H =
\left(\begin{matrix}
U+{\mu}&t_{12}&0&t_{14}\cr
t_{12}&U-{\mu} &t_{23}&0 \cr
0&t_{23}&U+{\mu} &t_{34} \cr
t_{14}&0&t_{34}&U-{\mu} \cr
\end{matrix} \right) .
\end{equation}
Choosing the parameter values 
\begin{align}
U=0.125,\;\;\;\; \mu=-0.55,\;\;\;\;t_{12}=t_{23}=t_{14}=t_{34}=-0.25, 
\label{param1}
\end{align}  
and diagonalizing this matrix, we find that the eigenvalues are given by
\begin{align*}
 E_1&= -0.6183034374  \\
 E_2&= -0.4249999999  \\
 E_3&=  0.6750000000  \\
 E_4&=  0.8683034374  
 \end{align*}
 The next step is to compute the spectrum from the Bethe Ansatz equations. Choosing 
 \begin{align}
\alpha_j&=\frac{1}{\sqrt{2}} \qquad j=1,2,3,4  
\label{alfas}
\end{align} 
 then (\ref{identify}) imposes the parameter values
\begin{align}
\eta = 1,\;\;\;\;\omega =1.1.\;\;\;\;\kappa = 0.5. 
\label{param2}
\end{align}
For the $N=1$ sector there are three pseudovacua.
\begin{itemize}
 \item  $|\phi_{0,0}\rangle= |0\rangle$: This pseudovacuum leads to the Bethe Ansatz equation
\begin{align*}
 \eta^2(v_1^2-\omega^2)=1 
\end{align*}
which has two solutions
\begin{align*}
v_1 = \pm 1.486606875
\end{align*}
Using (\ref{nrg}), this yields
\begin{align*}
 E(1.486606875) &= -0.6183034375 \\
 E(-1.486606875) &= 0.8683034375
\end{align*} 
in agreement with $E_1$  and $E_4$ obtained by numerical diagonalization.
 
 \item  $|\phi_{1,0}\rangle = \Gamma_{1,3}^\dagger|0\rangle$:
Here the pseudovacuum is an eigenvector 
of the Hamiltonian and through (\ref{nrg1}) we obtain
\begin{align*}
 E= -0.425
\end{align*}
which is in agreement with $E_2$ obtained by numerical diagonalization.

\item  $|\phi_{0,1}\rangle = \Gamma_{2,4}^\dagger|0\rangle$:
Here the pseudovacuum is an eigenvector 
of the Hamiltonian and through (\ref{nrg1}) we obtain
\begin{align*}
 E= 0.675
\end{align*}
which is in agreement with $E_3$ obtained by numerical diagonalization.

\end{itemize}

\subsection{$N = 2 $}
By using the normalized Fock basis  
$$\{|\chi_{i,j}\rangle=C_{ij}a_i^\dagger a_j^\dagger|0\rangle:  1\leq i\leq j =4, C_{jj}=1/\sqrt{2},\,C_{ij}=1 \mbox{ for } i\neq j\}$$ 
for the sector $N=2$, the Hamiltonian takes the matrix form
{\footnotesize
\begin{align*}
H=\left( \begin{matrix} 
{4U}+2\mu&{t_{12}}\sqrt{2}&0&0&0&0&{t_{14}}\sqrt {2}&0&0&0\cr
{t_{12}}\sqrt{2}&0&{t_{12}}\sqrt{2}&{t_{23}}&0&0&0&{t_{14}}&0&0\cr
0&{t_{12}}\sqrt{2}&{4U}-2\mu&0&{t_{23}}\sqrt {2}&0&0&0&0&0\cr
0&{t_{23}}&0&{4U}+2\mu&{t_{12}}&0&{t_{34}}&0&{t_{14}}&0\cr 
0&0&{t_{23}}\sqrt {2}&{t_{12}}&0&{t_{23}}\sqrt{2}&0&{t_{34}}&0&0\cr
0&0&0&0&{t_{23}}\sqrt {2}&{4U}+2\mu&0&0&{t_{34}}\sqrt {2}&0\cr
{t_{14}}\sqrt {2}&0&0&{t_{34}}&0&0&0&{t_{12}}&0&{t_{14}}\sqrt {2}\cr 
0&{t_{14}}&0&0&{t_{34}}&0&{t_{12}}&{4U}-2\mu&{t_{23}}&0\cr 
0&0&0&{t_{14}}&0&{t_{34}}\sqrt {2}&0&t_{23}&0&{t_{34}}\sqrt {2}\cr 
0&0&0&0&0&0&{t_{14}}\sqrt{2}&0&{t_{34}}\sqrt{2}&{4U}-2\mu  
\end{matrix} \right) 
\end{align*}
}

\noindent Again choosing the parameter values (\ref{param1}) to diagonalize the matrix we find that the eigenvalues are given by  
\begin{align*}
 E_1&= -1.12878  & E_6&= 0.233742  \\
 E_2&=  -0.883095 & E_7&= 0.2830951 \\
 E_3&=  -0.143398  & E_8&= 1.6000000   \\
 E_4&=  -0.600000 & E_9&=    1.743398\\
 E_5&=  0.0000000 & E_{10}&=  1.895046
 \end{align*}
To compare these results with those obtained from the Bethe Ansatz equations, we again make the identifications given by (\ref{alfas}) and (\ref{param2}). For this sector we have six pseudovacua.
 \begin{itemize}
  \item $|\phi_{0,0}\rangle= |0\rangle$: The pseudovacuum leads to the Bethe Anastz equations
\begin{align*}
 \eta^2(v_1^2-\omega^2)&=\frac{v_1-v_2-\eta}{v_1-v_2+\eta} \\
 \eta^2(v_2^2-\omega^2)&=\frac{v_2-v_1-\eta}{v_2-v_1+\eta}
\end{align*}
which admits the solutions
\begin{align*}
 &\quad \{v_1 = 1.128788719+0.4392638051i, v_2 = 1.128788719-0.4392638051i\},\\
 &\{v_1 = -1.779416015, v_2 = 1.311931196\}, \quad
 \{v_1 = -2.611876690, v_2 = -1.178215928\}
\end{align*}
as well as the spurious solutions sets 
\begin{align*}
\{v_1 = v_2=\pm 0.4582575695\} 
 \end{align*}
which must be discarded since the roots coincide (c.f. \cite{ik82}). Applying (\ref{nrg}) we find
\begin{align*}
E(1.128788719+0.4392638051i,1.128788719-0.4392638051i)&=  - 1.128788718  \\
E(-1.779416015,1.311931196)&= 0.2337424096
 \\
E(-2.611876690,-1.178215928)&= 1.895046310
\end{align*}
in agreement with $E_1,\,E_6$ and $E_{10}$ obtained by numerical diagonalization.

\item $|\phi_{1,0}\rangle= \Gamma_{1,3}^\dagger|0\rangle$: This pseudovacuum leads to the Bethe Ansatz equation
\begin{align*}
 \eta^2(v_1+\omega+\eta)(v_1-\omega) = 1. 
\end{align*}
The solutions are 
\begin{align*}
 v_1 &= 1.386796226411321,  -2.386796226411321
\end{align*}
 which produces the energy eigenvalues 
 \begin{align*}
 E(1.386796226411321) &=  -0.14339811,  \\
 E(-2.386796226411321) &=  1.74339811 
\end{align*}
agreeing with $E_3$ and $E_9$ obtained by numerical diagonalization.

\item $|\phi_{0,1}\rangle= \Gamma_{2,4}^\dagger|0\rangle$: Associated with this pseudovacuum there is the Bethe Ansatz equation
\begin{equation}
 \eta^2(v_1+\omega)(v_1-\omega +\eta) = 1 
\end{equation}
with solutions
\begin{align*}
 v_1 &= 0.666190378, -1.66619037. 
\end{align*}
These solutions provide the energy eigenvalues
 \begin{align*}
 E(0.666190378) &= -0.88309518   \\
 E(-1.666190378) &=  0.2830951894 
\end{align*}
corresponding to $E_2$ and $E_7$ obtained by numerical diagonalization.

\item $|\phi_{1,1}\rangle= \Gamma_{1,3}^\dagger \Gamma_{2,4}^\dagger |0\rangle $: 
Here the pseudovacuum is an eigenvector 
of the Hamiltonian and through (\ref{nrg1}) we obtain
\begin{align*}
 E= 0
\end{align*}
which is in agreement with $E_5$ obtained by numerical diagonalization.

\item $|\phi_{2,0}\rangle=\left(\Gamma_{1,3}^\dagger\right)^2 |0\rangle$:
Again the pseudovacuum is an eigenvector 
of the Hamiltonian and through (\ref{nrg1}) we obtain
\begin{align*}
 E= -0.6
\end{align*}
which is in agreement with $E_4$ obtained by numerical diagonalization.

\item $|\phi_{0,2}\rangle=\left(\Gamma_{2,4}^\dagger\right)^2 |0\rangle$:
Again the pseudovacuum is an eigenvector 
of the Hamiltonian and through (\ref{nrg1}) we obtain
\begin{align*}
 E= 1.6
\end{align*}
which is in agreement with $E_8$ obtained by numerical diagonalization.

 \end{itemize}
 
\subsection{Completeness of the Bethe Ansatz equations}
Our numerical investigations suggest that for $n=k+l<N$ fixed, there are $n+1$ sets of Bethe Ansatz equations (\ref{bae}) each admitting 
$N-n+1$ non-spurious solutions. Moreover,  when $n=N$ there are $N+1$ pseudovacuum states which are eigenstates of the Hamiltonian. This implies that the total number of Bethe states for fixed $N$ is
\begin{align*}
\sum_{n=0}^N (n+1)(N-n+1)=\frac{1}{6}(N+3)(N+2)(N+1)={\rm dim}(V_N). 
\end{align*}
Thus for generic values of the coupling parameters we expect that the Bethe Ansatz solution of this model is complete, analogous to other exactly solved models where this property has been established \cite{kl97,eks91,fk93,s94,flt99,b02}. 

\section{Conclusion}
Our main objective was to establish an exact Bethe Ansatz solution for the four-well Hamiltonian (\ref{h4}). This was achieved by first formulating the model through the Quantum Inverse Scattering Method. It was found that the model admits multiple pseudovacuum states, and for each of these there is a set of Bethe Ansatz equations given by (\ref{bae}).
Introducing new variables 
\begin{align*}
\tilde{\eta}&=(\alpha_1^2+\alpha_3^2)^{-1/4}(\alpha_2^2+\alpha_4^2)^{-1/4}\eta,  \\
\tilde{\omega}&=(\alpha_1^2+\alpha_3^2)^{-1/4}(\alpha_2^2+\alpha_4^2)^{-1/4}(\omega+\eta(k-l)/2), \\
\tilde{v}_j&=(\alpha_1^2+\alpha_3^2)^{-1/4}(\alpha_2^2+\alpha_4^2)^{-1/4}(v_j+\eta(k+l)/2)
\end{align*}
then (\ref{bae}) assumes the form 
\begin{align}
\tilde{\eta}^2(\tilde{v}_i+\tilde{\omega})(\tilde{v}_i-\tilde{\omega})
=\prod_{j\neq i}^{N-k-l}\frac{\tilde{v}_i-\tilde{v}_j-\tilde{\eta}}{\tilde{v}_i-\tilde{v}_j+\tilde{\eta}}.
\label{baenew} 
\end{align}
At this point we recognise (\ref{baenew}) as the Bethe Ansatz equations for the two-mode Bose-Hubbard model \cite{esks91,esse92,eks93}. 
It turns out that the eigenspectrum of the four-well Hamiltonian (\ref{h4}) is in one-to-one correspondence with the eigenspectrum of a particular {\it ensemble} of two-well Hamiltonians, with particle numbers and coupling parameters dependent on the eigenvalues $k$ and $l$ of the additional conserved operators  as given by (\ref{cons1}) and (\ref{cons2}). This opens a path of investigation into the four-well model by utilizing results known for the two-well model. This approach will be developed in future work, with particular focus on characterizing the ground-state properties of the four-well model.

 \vspace{1.0cm}
\centerline{{\bf Acknowledgements}}
~\\
Jon Links and Angela Foerster are supported by the Australian Research Council through Discovery Project DP150101294.

\end{document}